\documentclass[twocolumn,aps,floats,floatfix,superscriptaddress]{revtex4}
\usepackage{graphicx,amssymb}
\usepackage{epsfig}
\newcommand{\beq}{\begin{equation}}
\newcommand{\eeq}{\end{equation}}

\newcommand{\br}{{\bf r}}

\newcommand{\bR}{{\bf R}}
\newcommand{\bG}{{\bf G}}
\newcommand{\bK}{{\bf K}}

\newcommand{\bq}{{\bf q}}

\begin{document}
\draft
\title{Exciton solid in bilayer two dimensional electron-hole systems}
\author{ S. T. Chui}
\affiliation{ Bartol Research Institute and Dept. of Physics and Astronomy, 
University of Delaware, Newark, DE 19716, USA}
\author{Ning Wang}%phwang@ust.hk
\affiliation{ Dept of Physics, Hong Kong University of Science and Technology, Clear Water Bay, Hong Kong}
\author{Chun Yu Wan} % <cywanag@connect.ust.hk}
\affiliation{ Dept of Physics, Hong Kong University of Science and Technology, Clear Water Bay, Hong Kong}

%\date{\today}
\begin{abstract}
We propose a state of excitonic solid for double layer two dimensional electron hole systems in transition metal dicalcogenides
stacked on opposite sides of thin layers of BN. Properties of the exciton lattice such as its Lindemann ratio and possible supersolid behaviour are studied. We found that the solid can be stabilized relative to the fluid by the potential due to the BN.
\end{abstract}
\maketitle
\section{Introduction}
There has been much recent interest recently in bilayer electron-electron (e-e) and electron-hole (e-h)  in graphene and transition metal dicalcogenides (TMDC) systems.
This follows earlier interest in the physics of the two dimensional electron gas
 in single and double layers in Si-MOSFET and in GaAs heterostructures. For a single layer, the electrons are expected to be a fluid at high densities and a solid at low densities\cite{chui}. 
For the bilayer electron-hole system, there has been much interest in the possibility that the 
electron and hole forming an exciton which then  Bose condense into a superfluid.
This is usually expected in the limit with the exciton size less than the average interparicle spacing so that the identity of the exciton is well defined. 
The possibility of a Bose condensate of excitons in this system has recently been discussed by Wang and coworkers\cite{Mak}. Optical evidence is presented for the existence of excitons but no transport measurement have been carried out. As far as we know, there has been no study on the possibility of the excitons forming a solid instead of a Bose fluid. A boson can in principle exhibit both phases, as is exemplified by the example $^4$He, which can exist in both a solid and a superfluid phase.  The energetics of this exciton solid is different from that of the Wigner crystal, which is predicted to be stable at low densities.  At large distances, the excitons interact with each other with dipolar interactions, much weaker than the Coulomb interaction of the electron solid. The exciton solid by itself is not stable at low densities. In this paper we study the physical properties of the exciton solid. 

Monolayer graphene exhibit a linear dispersion near the Fermi energy and the physics is different from those  with parabolic particle bands, which are exhibied by two classes of experimental systems
under active study at the moment. These are double layer graphene and double layer TMDC system sandwiching a BN layer.
The effective masses of the particles in the double layer graphene ($0.03 m_e$) is much smaller than that in the TMDC ($0.7 m_e$.) 
The thin BN layer is of thickness $d_0\approx 5 nm$ so that the electrons and holes
in layers on opposite sides of BN have a separation larger than $d_0$.   Following our previous work on the phonons of rare gases adsorbed on graphite\cite{Toufic}, we carried out self-consistent phonon calculation of the excitonic solid system. A criterion of the stability of the solid is the Lindemann ratio, usually defined as the root mean square lattice vibration normalized by the lattice constant. In two dimension the root mean square vibration can be infinite. A useful alternative is to focus on the relative vibration between nearest neighbours. For quantum melting, the melting point occurs when this ratio is about 10 per cent\cite{chui}.
We found that for current experimental parameters, the Lindemann ratio for TMDC is about 20 per cent; for double layer graphene, 63 per cent. However, the excitons are in the presence of the BN layer. There is charge transfer between Boron and Nitrogen of about $Q\approx 0.47$ in BN\cite{x}. We study the phonon in the presence of the external potential due to BN and show the Lindemann ratio can be changed.  We  found that the Lindemann ratio is reduced to 7 per cent for TMDC and 27 per cent for bilayer graphene. This suggests that such an excitonic solid state is indeed possible for current experimental system with TMDC layers.
Supersolid behaviour has been discussed for Boson solids such as solid $^4$He. We
discuss the possible manifestation of this kind of behaviour for the current system. 
%We discuss the transport properties of this state and found that it is %consistent with recent results of bilayer double graphene\cite{UT} with a large %negative drag resistance when the number of electrons is equal to the number of %holes. This state occurs when the number of electrons is equal to the number of %holes The drag resistance of this state  
\section{Exciton in the double layer structure}
To gain some intuition of the properties of the system, we first discuss the  physical property of a single exciton in the bilayer structure with BN in between.
We assume an average dielectric constant of 4 from the screening by the BN layer and an effective mass of 0.7 the electron bass. The effective mass $\mu$ for the exciton is twice the electron mass. The typical size of an "ordinary" exciton is two times the Bohr radius.
The Bohr radius for the graphene system ($a_{B, graphene}=90\AA$) is usually larger than the effective BN layer thickness film that is of the order of $d=d_0(\epsilon_{xy}/\epsilon_z)^{1/2}=d_0(6.93/3.76)^{1/2}\approx 67.9\AA$ . The Bohr radius  for the TMDC system $a_{B, TMDC}=3\AA$. This exciton size is smaller than the thickness of the BN layer, however. 
The exciton binding energy for the double layer graphene system is of the order ${\rm Ryberg}\ \mu/m_e/\epsilon^2=137K$ .
For double layer TMDC, the conventional formula gives a binding energy more than twenty times larger.
It is much more difficult to ionize them. 
%For bl graphene, $Ryb=13.6*0.03/16 *10600 K=273 K.$ With the effective mass
%$m=m^2/2m=m/2.$ The binding energy is even less. 
However, the exciton in the TMDC system needs to be reexamined since the Bohr radius is now less than the BN layer thickness. The motion of the electron hole system is now two dimensional. We discuss this next.

The transverse separation $r$ of the two dimensional exciton is at most of the order of the Bohr radius between the the electron and the hole in the TMDC and is much less than their vertical spacing $d$. The Coulomb potential energy between the electron and the hole can be expanded as a power series in $r/d$ as $$V=-e^2(r^2+d^2)^{-1/2}
%\approx -e^2 (1/d-r^2/d^3/2) 
\approx -e^2 /d+e^2 r^2/d^3/2.$$ $e^2=q_e^2/(4\pi\epsilon).$ In this approximation, the exciton
wavefunction is that of a two dimensional harmonic oscillator  with a force constant $k=e^2/d^3=\mu\omega^2$ and $\omega=[e^2/d^3/\mu]^{1/2}.$
%$\hbar \omega=e^2/d [\hbar^2 /e^2\mu]^{1/2} [1/d]^{1/2}=e^2/d [a_B/d]^{1/2}.$
The exciton binding energy is $E_{ex}=-e^2/d+\hbar\omega/2$. This can be written as
\beq
E_{ex}=e^2/a_B[-a_B/d +(a_B/d)^{3/2}/2]
\eeq
%For double layer graphene , with the bare Bohr radius $a_{B0}=0.54\AA,$
%For graphene $a_B=(\epsilon/\epsilon_0)m/m^*a_{B0}=4 a_{B0}/(0.03)\approx 133 a_{B0}\approx %70.4\AA.$
%For TNDC $a_B=(\epsilon/\epsilon_0)m/m^*a_{B0}=4 a_{B0}/(0.7)\approx 6 a_{B0}\approx %3 \AA.$
For TMDC 
%$e^2/a_B\approx e^2/(6a_{B0}) \approx 13.6\times 11600 K /6\approx 589 K.$
with $d=68\AA$, $\epsilon=4,$ 
%$e^2/d\approx e^2/(2a_{B0})(2a_{B0}/d)\approx 13.6 ev/67/\epsilon\approx %13.6\times 116 K /2.68\approx 589 K.$
%$E_{ex}=394 K([-1+(70/d)^{1/2}]$. It is very easy to ionize them.
%$(a_B/d)^{1/2}=(14/68)^{1/2}=0.45,$ 
$E_{ex}=260 K.$
%The extremum exciton energy occurs when $+1/d^2-a_B^{1/2}3/2d^{-5/2}2=0.$
%THis implies $a_B (9/16)= d$ at which our approximation in invalid. No extremum.

%For  TMDC,  %$a_B\approx 3.5$ $a_B/d<<1.$  
The size of the exciton is $\xi=[\hbar/(\mu\omega)]^{1/2}
%=[\hbar d^{3/2}/[\mu e^2]^{1/2} ]^{1/2}.$ This can be written in terms of the %Bohr radius $a_B=\hbar^2/(\mu e^2)$ as $\xi=(a_B d^3)^{1/4}.$
%Thus $\xi/ d 
\approx  (a_B/d)^{1/4}\approx 1/2.$ For current experimental parameters,  $\xi\approx 25\AA$ is less than the average interparticle distance that is of the order of 100\AA. Thus the condition of well defined excitons is satisfied. 
%$r_s=a_0/a_B,$ $a_0=(\pi n)^{-1/2}.$ For a triangular lattice with lattice %spacing $a=76\AA$, $a_0=a(\sqrt{3}/2/\pi)^{1/2}=0.525 a.$ 
%In this paper we $r_s=0.525 a/3=13.3.$
%$n=1/(a^2\sqrt{3}/2)=10^{12}/cm^2/(0.76^2*0.866)=2\times 10^{12}/cm^2.$
The electroluminence is proportional to the probability of finding the electron and hole on top of each other and thus equal to $\psi_{exciton}(r=0)\propto 1/\xi^2$.
\section{Exciton solid Phonons: Optical and Acoustic mode}
\begin{figure}[tbph]
%\vspace*{0pt} \centerline{\includegraphics[angle=0,width=6cm]{itd1}} 
%\vspace*{0pt} \centerline{\includegraphics[angle=0,width=6cm]{scp/%phononcm1.eps}} 
\vspace*{0pt} \centerline{\includegraphics[angle=0,width=6cm]{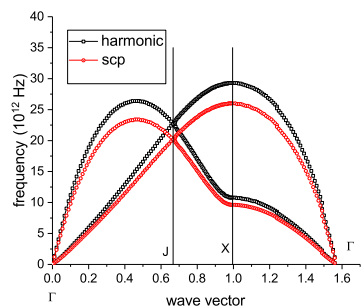}} 
\vspace*{0pt}
\caption{ 
Phonon energy for the center of mass along symmetry directions of the two dimensional Brillouin zone for $r_s=13.3$, corresponding to a density of $ 2\times 10^{12}/cm^2$.}
\label{phonon1}
\end{figure}

We next examine the phonon excitations when the  excitons are arranged in a lattice. We ignore the motion of the particles perpendicular to the layer and focus on their in plane motion.  The lattice motion of the exciton solid consists of two modes, an optic mode from the intra-exciton vibration and an acoustic mode from the inter-exciton vibration. We separate the optical mode  from the acoustic mode as follows. We assume the lattice positions to be at $\bR_i$.
At each site, we write the in plane coordinates of the electrons and holes 
in terms of the center of mass equilibrium and fluctuation positions ($\bR_i+\delta \bR$) and the relative ($\br_i$) coordinates so that the coordinate of the i$^{th}$ electron is given by $\br_{ei}=\bR_i+\delta \bR_i+\br_i/2;$ of the hole, $\br_{hi}=\bR_i+\delta \bR_i-\br_i/2.$ The potential energy of interaction involves $U=\sum_{\alpha,\beta=e,h,i,j} V[(r_{\alpha,i}-r_{\beta,j})^2+d^2\delta_{\alpha,\beta}]^{1/2}/2$ where $V(u)=e^2/u$ is the Coulomb potential. We expand the interaction in powers of $\delta R_i$  and $r_i$. Since $U$ is even in $r_i$, only even powers of this quantity appears in a power series expansion. Thus in the phonon approximation when the second order terms are kept
we get $U=\sum_{i\neq j} (\delta R_i-\delta R_j)^2\cdot\nabla\nabla V_{ij}+\sum_{i,j}(r_i-r_j)^2\cdot \nabla\nabla V_{ij}.$ The two degrees of freedom are decoupled. For the relative coordinate, there is an additional intra-exciton interaction for $i=j$ in the sum for $U$.
Following our previous work on the phonons of rare gases adsorbed on graphite\cite{Toufic}, we carried out self-consistent phonon calculation of this bilayer system. The harmonic frequency is obtained from solving the phonon equation for the Fourier transform phonon modes: 
\beq
m\omega^2 e_i(q)=\sum_j e_j(q) D_{ij}(q)
\eeq
where the dynamical matrix $D_{ab}(q)=\sum_{\bR}(1-\cos \bq\cdot \bR)\nabla_a\nabla_b U(R)$, $e_i(q)$ is the polarization vector.
The harmonic and the self consistent acoustic center of mass phonon frequencies for an triangular exciton lattice for parameters of our recent experiments ( density $2\times 10^{12}/cm^2$
lattice spacing 71\AA ) are shown in Fig.\ref{phonon1} as a function of the wave vector along symmetry directions.
Also shown in this figure are the self consistent phone frequencies obtained 
by replacing the dynamical matrix $D$ by an average $<D>$ so that
$<D_{ab}(q)>=\sum_{\bR}(1-\cos \bq\cdot \bR)\nabla_a\nabla_b <U(R+\delta R)>$.
The angular brackets indicate a self-consistent thermal and quantum average over the fluctuation $\delta R.$ 

\begin{figure}[tbph]
%\vspace*{0pt} \centerline{\includegraphics[angle=0,width=6cm]{itd1}} 
%\vspace*{0pt} \centerline{\includegraphics[angle=0,width=6cm]{scp/%phononcmo1.eps}} 
\vspace*{0pt} \centerline{\includegraphics[angle=0,width=6cm]{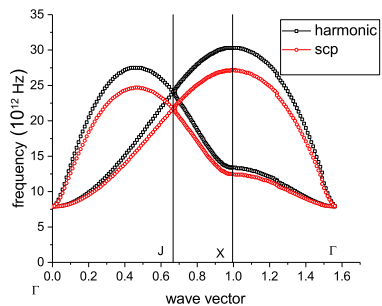}} 
\vspace*{0pt}
\caption{ 
Phonon energy for the intraexciton  mode along symmetry directions of the two dimensional Brillouin zone.}
\label{phonon1o}
\end{figure}

\begin{figure}[tbph]
%\vspace*{0pt} \centerline{\includegraphics[angle=0,width=6cm]{scp/phononcm2.eps}} 
\vspace*{0pt} \centerline{\includegraphics[angle=0,width=6cm]{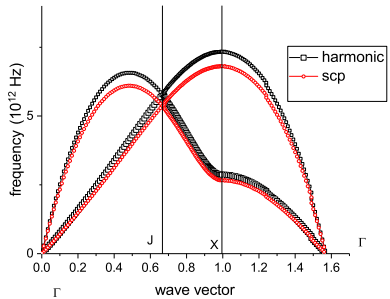}} 
\vspace*{0pt}
\caption{ Phonon energy for the acoutic mode along symmetry directions of the two dimensional Brillouin zone. The lattice constant is twice that in the previous two figures with $r_s=26.6$, corresponding to a density of $0.5\times 10^{12}/cm^2$.}
\label{phonon2}
\end{figure}

The harmonic and the self consistent optic exciton vibration frequencies are shown in Fig.\ref{phonon1o} as a function of the wave vector along symmetry directions.
We have explored the dependence of this on the density of the system. The harmonic and the self consistent acoustic center of mass phonon frequencies for an triangular exciton lattice of a larger spacing of spacing 152\AA\  are shown in Fig. \ref{phonon2} as a function of the wave vector along symmetry directions. 
\section{Lindemann ratio}
We discuss first the qualitative behaviour of the Lindemann ratio for our system.
The interparticle potential has two regimes of behaviour. At small (large) distances with $r<<d$ ( $r>>d$), $V\approx e^2\alpha r^{-n}$ where $n=1$, $\alpha=1$
($n=3$, $\alpha=1/d^2$).
The mean square lattice vibration is $<r^2>=\sum_q \hbar(2n_q+1)/(m\omega_q)$ where $n_q$ is the number of phonons. The Bohr radius is $a_B=\hbar^2/(me^2).$ For a general interparticle potential $V\approx e^2\alpha r^{-n}$ the phonon frequency is given by
$m\omega^2= V''=e^2\alpha a^{-n-2}.$ 
$\omega=e(\alpha/m)^{1/2}a^{-1-n/2}.$ The Lindemann ratio at low temperatures is given by
$<r^2>/a^2=(a_B/\alpha)^{1/2}a^{n/2-1}.$
We thus expect that at very high densities with small $a<d$, 
$<r^2>/a^2=(a_B/a)^{1/2}.$ 
Correspondingly at very low densities with large $a>d$, $<r^2>/a^2=(a_Ba)^{1/2}/d.$ 
For both very large and very small a, the fluctuation and the Lindemann ratio is large;
one has a fluid. This is in contrast with the Wigner crystal (n=1) where at large a, the Lindemann ratio, $(a_B/a)^{1/2}$, is small and the solid is stable.
%The crucial quantity is $(a_B/d)^{1/2}.$ In the present case, at low densities, the interexciton potential is a dipolar interaction and much weaker. 
\begin{figure}[tbph]
%\vspace*{0pt} \centerline{\includegraphics[angle=0,width=6cm]{scp/lmrn.eps}} 
\vspace*{0pt} \centerline{\includegraphics[angle=0,width=6cm]{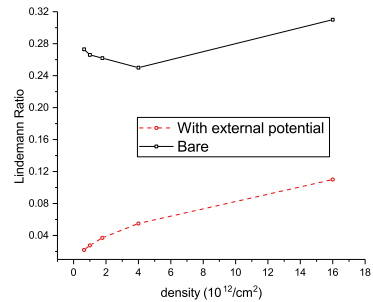}} 
%\vspace*{0pt} \centerline{\includegraphics[angle=0,width=6cm]{scp/lmr.eps}} 
%\vspace*{0pt} \centerline{\includegraphics[angle=0,width=6cm]{scp/lrmic.eps}} \vspace*{0pt}
\caption{ The Lindemann ratio as a function of density with and without the external potential from BN.}
\label{lmr}
\end{figure}

Our numerical result for the Lindemann ratio is shown in Fig. (\ref{lmr}) as a function  of $r_s$. It is around 20 per cent and not a strong function of the density, as is indicated above.
\section{Effect of BN}
In the two dimensional electron system, the external static potential such as that due to the dopants can stabilize the solid phase relative to the fluid phase\cite{chuiimp}. In the current system, there is a thin boron nitride film in the middle in the experimental systems. The two dimensional electrons and holes see a static potential from the periodic array of boron and nitrogen ions. 
%This periodic potential can enhance the stability of the solid phase. 
We find this potential produces a significant effect in reducing the spatial fluctuation of the exciton solid and enhance its stability. We calculate this next.

There is a charge transfer between boron and nitrogen in boron nitride of magnitude\cite{x} $Q\approx 0.47$.
We calculate this potential by decomposing the potential from BN film as consisting of a sum of two dimensional periodic arrays at different vertical distances $h$ away from the two dimensional electron (hole) gas.  
The potential of each two dimensional periodic array is sum of the Coulomb potential $U$ from the ions at the two dimensional periodic BN lattice sites $\bR^{B,N}_i+\delta \bR^{B,N}_i$ with modulation $\delta \bR_i^{B,N}=\sum_G \delta \bR_G^{B,N} e^{i\bG\cdot\bR_i}$ from the interaction of the BN latice 
with the incommensurate TMDC lattice with reciprocal lattice vectors $\bG$.  
We assume that the hole lattice is displaced  by  a slight amount relative to the electron lattice so that both the electron and the hole lattices are close to the minima of the periodic potential\cite{detail}. 

%For example, t
The potential from the boron ions is given by $V_B(r)=\sum_i U_B(r-\bR^{B}_i-\delta\bR^{B})$. This sum can be written in Fourier space as $V_B(r)=\int d^2q /(2\pi)^2 \sum_i {\tilde U}_qe^{i(r-\bR^{B}_i-\delta\bR^{B})}$ where the Fourier transform of the Coulomb potential ${\tilde U}(q)=\int d\br e^{-iq\cdot r}U(r)=Q \sum_h e^{-|q|h}/|q|/\epsilon/2$  Because the boron and the nitrogen ions have opposite charges, we assume $\delta \bR^N_G=-\delta \bR^B_G.$ To illustrate our result we have used a typically small strain of $\delta \bR^N_G/a_{BN}$ of 2.5 per cent.
%From Coulomb's law
%$\nabla^2 U(r)=-\rho/\epsilon.$ Let $U(r)=\int d\bq e^{iqr}U(q)/(2\pi)^3.$
%For a point source $\rho(r)=\int d\bq {\tilde \rho}(q)e^{iq\cdot r},$
% ${\tilde \rho(q)}=Q/(2\pi)^3.$ We get 
%$\nabla^2 \int d\bq e^{iqr}U(q)/(2\pi)^3=-\int d\bq {\tilde \rho}(q)e^{iq\cdot %r}/\epsilon.$
%$-(q^2+q_z^2)U(q)/(2\pi)^3=-Q/(2\pi)^3.$
% $U(q)=1/(q^2+q_z^2)/\epsilon.$
%$$U(r,z)=Q \int d^2qdq_z e^{i(q_z (h-z)+q\cdot (r-r_0))}/(q^2+q_z^2)/\epsilon/%(2\pi)^3 $$
%$$=Q \pi\int d^2q e^{-q|h-z|+iq\cdot (r-r_0)}/q/\epsilon/(2\pi)^3.$$
%${\tilde U}(q) 
%=\int d\br e^{-iq\cdot r}U(r,z=0)

In terms of the reciprocal lattice vectors $K$ of the BN lattice
%as $V_B(r)=\sum_K {\tilde U}(K) e^{iK\cdot r}/a_c,$ 
$$
V_B(r)\approx  \sum_K[V_{B1}^K(r)+\sum_G V_{B2}^{K-G}(r)]$$
$V_{X1}^K(r)={\tilde U}_Ke^{i\bK\cdot \br}/a_c,$ $V_{X2}^{K-G}(r)=
i {\tilde U}_{G-K}e^{i(\bG-\bK)\cdot \br} [(\bG-\bK)\cdot\delta\bR^X_{G})]/a_c;$
$a_c$ is the area of the unit cell of the BN lattice.
%V_B(r)=\sum_{K,h} e^{iK\cdot r}Q e^{-|K|h}/|K|/\epsilon/2/a_c.
Because of the factor $e^{-|K|h}$ $V_B$ is dominated by contributions from the nearest two BN planes with $h_1\approx 3.5 \AA$, $h_2\approx 7 \AA$ and from the smallest reciprocal lattice vector $K=2\pi/a_{BN}$ and $K-G=2\pi(1/a_{BN}-1/a_{TMDC}).$
%To be checked: factors of $\pi.$

\begin{figure}[tbph]
%\vspace*{0pt} \centerline{\includegraphics[angle=0,width=6cm]{scp/phononex1.eps}} 
%\vspace*{0pt} \centerline{\includegraphics[angle=0,width=6cm]{scp/%dispic5.eps}} \vspace*{0pt}
\vspace*{0pt} \centerline{\includegraphics[angle=0,width=6cm]{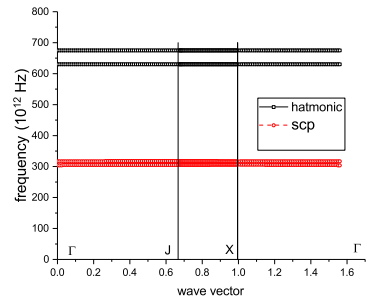}} \vspace*{0pt}
\caption{ Phonon energy for the center of mass along symmetry directions of the two dimensional Brillouin zone in the presence of the external potential due to BN for a density of $2\times 10^{12}/cm^2$ and $\delta R_G/a_{BN}=0.025$. }
\label{phononex1}
\end{figure}

The averages of the pinning potentials over the spatial fluctuations of the excitons are determined by the corresponding Debye-Waller factors as:
\beq
<V_{Bi}^Q(r)>\approx  V_{Bi}^Q e^{-<(Q\cdot \delta r)^2>/2}
%\sum_i {\tilde U}_K[e^{iK\cdot r-K^2<(\delta r)^2/2}
%+i \sum_G {\tilde U}_{G-K}[e^{i(G-K)r-(G-K)^2<(\delta r)^2/2} J_1((G-K)\delta%\bR^{G})/a_c.
\eeq
In our calculation, this effect is determined self-consistently. We find that the Debye-Waller factor for $V_{B1}$ is extremely small because $<(K\cdot \delta r)^2>$ is large. The average pinning potential $V_{B2}$ with a wave vector $|G-K|\approx |K|/4<|K|$ is much bigger.
Similarly the potential from the nitrogen ions at sublattice position $\delta$ is given by $V_N=-\sum_i U(r-R_i-\delta R_i^N-\delta)$ where the boron nitrogen distance is $\delta=0.1446 nm$. This can be written as a Fourier series:
\beq
V_N(r)\approx  -\sum_K[V_{N1}^K(r)e^{-iK\cdot\delta}+\sum_G V_{N2}^{K-G}(r)e^{-i(K-G)\cdot\delta}]
\eeq 
%\sum_i {\tilde U}_K[e^{iK\cdot (r-\delta)}
%+i \sum_G {\tilde U}_{G-K}[e^{i(G-K)\cdot(r-\delta)} J_1((G-K)\delta\bR^{G})/%a_c.
%V_N=-\sum_{K,h} e^{iK\cdot (r-\delta)} Q e^{-|K|h}/|K|/\epsilon/2/a_c
%e^{-|q|h+iq\delta}/|q|/\epsilon/2.
The BN potential is incommensurate with the lattice constant of the exciton solid; misfit dislocations will form\cite{fv,ying}. 
The lattice constant $a$ of the exciton solid lattice is very different from that of BN potential so that $\gamma=a(G-K)/(2\pi)$ is big. We approximate $\gamma$ by the closest integer $\gamma_0$. Over most [$1-(\gamma-\gamma_0)/\gamma$] of the Brillouin zone away from the zone center, the phonons are well approximated by that of a commensurate one with periodicity $a_0=2\pi \gamma_0 / (G-k)$\cite{ying} so that $(a_0-a)/a$,   the difference in periodicty between the commensurate and the incommensurate case, is small. In the small region close to the zone center, $|q|<[1-(\gamma-\gamma_0)/\gamma]\pi/a$ the phonons are gapless and correspond to the movement of the misfit dislocations. An example of the approximate phonon dispersion for the center of mass in the presence of this potential away from the zone center is shown in Fig. (\ref{phononex1}) for a density with $r_s=13.3$ and $\delta R^B
_G/a_{BN}=0.025.$ In this figure, we show both the harmonic frequency without the Debye Waller factor and the self consistent phonon (scp) frequency where the Debye Waller factor is self consistently determined. The contribution from $V_{BN,1}$ is significant in the harmonic approximation and becomes very small in the scp result.

%The electron and hole have opposite charges. To take advantage of the BN %potential, we expect the exciton to be distorted laterally by a distance $%\delta.$. 

Relative to the bare phonon frequency, the phonon frequency with the pinning potential from BN is increased and the lattice vibration amplitude is reduced. The Lindemann ratio for the exciton solid in the presence of this potential is shown in Fig.(\ref{lmr}) as a function of the density of the system. This factor is now less than around ten per cent and suggests that the solid can be stabilized by the BN potential.
\begin{figure}[tbph]
%\vspace*{0pt} \centerline{\includegraphics[angle=0,width=6cm]{scp/lmrt.eps}} 
\vspace*{0pt} \centerline{\includegraphics[angle=0,width=6cm]{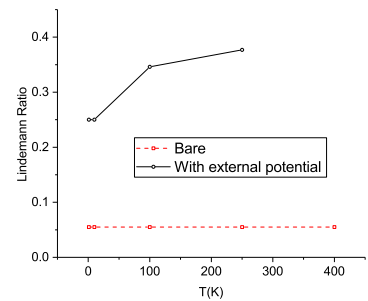}} 
%\vspace*{0pt} \centerline{\includegraphics[angle=0,width=6cm]{scp/lrmtic.eps}} \vspace*{0pt}
\caption{ The Lindemann ratio as a function of temperature
a density of $4\times 10^{12}/cm^2$ with and without the external potential from BN.}
\label{lmrt}
\end{figure}
The temperature dependence of this ratio is shown in Fig \ref{lmrt} 
a density of $4\times 10^{12}/cm^2$. This suggests that the solid is stable at low temperatures .

\section{Possible "supersolid" behaviour}
There has been much recent interest in the supersolid behaviour of Boson solids such as $^4$He. This kind of phenomenon is believed to come from the quantum transport of defects. In our system the defect energy is lower at the edge so we expect the density of defects to be higher there. Thus we expect
the possible "supersolid" behaviour to be dominated by the one dimensional quantum transport of defects at the edges. There has been much interest in the quantum transport of electrons in wires\cite{ibm}, which manifested in a conductance of the order of $G_0=2e^2/h.$ In one dimension there is no difference in statistics between impenetrable Bosons and Fermions\cite{Giradeau}. 
We thus expect a similar conductance to be manifested in the Coulomb drag measurement, which we have recently observed\cite{Wang} and which motivated the   
current work.
\section{Conclusion}
In conclusion we examine a collection of excitons for the limit that the exciton size is less than the exciton separation so that the excitons are not interpenetrating. We
propose that a collection of these excitons can also form a solid in addition to being a fluid. The stability of this in TMDC bilayer systems is investigated. The external potential from the BN substrate is found to be important and can stabilize the solid phase. Our description for the exciton solid differs from that
for the other limit when the excitons size is larger than the exciton separation. In that case the excitons are interpenetrating and their identity is not well defined.
This other limit corresponds to the charge density wave\cite{cdw}/excitonic insulator\cite{HR}, which is usually described by a generalization of the mean field BCS type wave function $\Pi (u_k-v_kb^+_{k+Q}a_k)|0>$ for electron (hole) operators $b_p^+$ ($a_q$) and coefficients $u$ and $v$. In real space this function $\Psi$ is a "Slater determinant" of particle hole functions $f(r_e-r_h)$ with specific two body momentum correlations\cite{dyson}: $\Psi=\sum_{P,Q} (-1)^{P+Q}\Pi_{i,j}f(r_{ePi}-r_{hQj})$ where we sum over all permuatations P and Q of the particle and hole indices. The order parameter in this limit is $<\rho_q>=<\sum_p b^+_{p+q}a_p>$, the average charge density of specific momentum $q$. The stability of this phase comes from the nesting of Fermi surfaces of particle and hole bands with momentum difference q  so that the kinetic energy cost of forming an exciton is particularly small. This nesting of Fermi surfaces is absent in the bilayer TMDC system.

The electron hole system can also exist in a metallic state of an electron-hole plasma, as has been observed in bulk Ge under laser excitation\cite{ehp, BRAC}.
We have explored such a state for the current system with fixed node quantum Monte Carlo simulation\cite{qm} and found that for a single valley spin polarized system, the metallic state is more favorable for $r_s<5$ for the bilayer system. This is more likely to be observed in bilayer graphene. For the similar density, the corresponding $r_s$ is much smaller because of a much higher Bohr radius in the graphene system.
\acknowledgements{N.Wang thanks the support from the 
Research Grants Council of Hong Kong (Project No. 16300717 )}.


\begin{references}
\bibitem{detail}
For a displacement $\delta a$ The energy cost is  $e^2 (\delta a)^2/d^3\approx (\delta a/d)^2 e^2/d\approx 4\times 10^{-4} e^2/d.$ This is much smaller than the pinning potential.
\end{references}
\end{document}